\documentclass{vgtc}
\pdfoutput=1
\usepackage{mathptmx}
\usepackage{graphicx}
\usepackage{times}
\usepackage{url}
\usepackage{subfigure}
\usepackage{fancyvrb}
\usepackage[center]{caption}

\newcommand{\Tool}{\textsc{Zifazah}}

\title{Composing DTI Visualizations with End-user Programming}

\author{Haipeng Cai, Jian Chen, Alexander P. Auchus, and David H. Laidlaw}
\authorfooter{
\item
 Haipeng Cai is with University of Southern Mississippi, e-mail: haipeng.cai@eagles.usm.edu.
\item
 Jian Chen is with University of Southern Mississippi, e-mail: jian.chen@usm.edu.
\item
 Alexander P. Auchus is with University of Mississippi Medical Center, e-mail: aauchus@umc.edu.
\item
 David H. Laidlaw is with Brown University, e-mail: dhl@cs.brown.edu.
}

\shortauthortitle{Haipeng Cai\MakeLowercase{\textit{et al.}}: {\Tool}}

\abstract{
We present the design and prototype implementation of a scientific visualization language called {\Tool} for composing
3D visualizations of diffusion tensor magnetic resonance imaging (DT-MRI or DTI) data. Unlike existing tools allowing
flexible customization of data visualizations that are programmer-oriented, we focus on domain scientists as end users in order to enable them to freely compose visualizations of their scientific data set. We analyzed end-user descriptions extracted from interviews with neurologists and physicians conducting clinical practices using DTI about how they would build and use DTI visualizations to collect syntax and semantics for the language design, and have discovered the elements and structure of the proposed language. {\Tool} makes use of the initial set of lexical terms and semantics to provide a declarative language in the spirit of intuitive syntax and usage. This work contributes three, among others, main design principles for scientific visualization language design as well as a practice of such language for DTI visualization with {\Tool}. First, {\Tool} incorporated visual symbolic mapping based on color, size and shape, which is a sub-set of Bertin's taxonomy migrated to scientific visualizations. Second, {\Tool} is defined as a spatial language whereby lexical representation of spatial relationship for 3D object visualization and manipulations, which is characteristic of scientific data, can be programmed. Third, built on top of Bertin's semiology, flexible data encoding specifically for scientific visualizations is integrated in our language in order to allow end users to achieve optimal visual composition at their best. Along with sample scripts representative of our language design features, some new DTI visualizations as the running
results created by end users using the novel visualization language have also been presented.
}


\CCScatlist{ 
 \CCScat{H.5.2}{User Interfaces}%
{Programming Language};
 \CCScat{I.3.6}{Methodology and Techniques}{Languages}
}

\begin{document}

\firstsection{Introduction}
\label{sec:intro}

\maketitle

Visualization tools often support user customization, which allows changes of the visualization so
as to help users gain better understanding of the underlying data thus to facilitate knowledge discoveries about the data that would be hard to achieve otherwise. However, the support of user creativity is usually constrained by the limits of
predefined options or functionalities for the customizations. 

An effective way to address these constraints is to
offer users a programming environment in which they can freely compose towards desirable visualizations of their data
through a visualization language. While such languages have been proposed and successful in the information visualization
(InfoVis) community ~\cite{fry2004computational, mackinlay2007show, bostock2009protovis}, there is a lack of
end-user visualization language for 3D scientific visualization (SciVis). Based on our many discussions with domain users, we have recognized that domain scientists want a visualization of their own data to be designed and built by themselves. Now that the success of visualization languages for InfoVis is probably attributed to their capabilities of empowering users to design their own visualizations, what if domain scientists have a visualization tool that is powerful but easy to maneuver so that they can fully control the design elements and visual components to create whatever visualization they really want in mind?

A recent advanced MRI technique, diffusion tensor imaging (DTI) has proven advantageous over other imaging techniques
in that it enables \textit{in vivo} investigation of biological tissues and, through three-dimensional (3D) tractography
~\cite{Basser-2000-IFT}, explorations of the distribution and connectivity of neural pathways in fibrous tissues like brain whiter matter and muscles. Further, as one way to visualize DTI data, 3D visualization of the streamline data model derived from the tractography can illustrate the connectivity of fiber tracts and structures of anatomy, and therefore provides a powerful means that assists neuroscientists in clinical diagnosis and neurosurgical planning.

We proposed a visualization language as the first tool of this kind for DTI visualizations because DTI is complex enough
to stimulate a design that would be useful for simpler and similar visualization problems such as that of flow
visualization. Although mainly driven by neurologists' need for conducting their clinical tasks
with DTI visualizations, our language design would also be reusable in a broader range in 3D SciVis.
Motivated by the needs of spatial explorations in 3D scientific visualizations because of the spatial constraints within
the data, the present language is particularly useful in empowering domain scientists to build 3D visualizations that best meet their specific needs.

Furthermore, the language can facilitate domain scientists' effective use and exploration of the visualizations as well, because it allows them to customize essential elements of visualization with the maximal flexibility by applying their best understanding of the domain data to the visualization composition process. Illuminated by Bertin's \textit{Semiology of Graphics} ~\cite{bertin1983semiology}, we design the language to allow users to compose symbols in 3D visualizations, including visual encoding methods and other causes that affect visualization task performance.
\begin{figure}[htb]
\centering
\includegraphics[width=8cm]{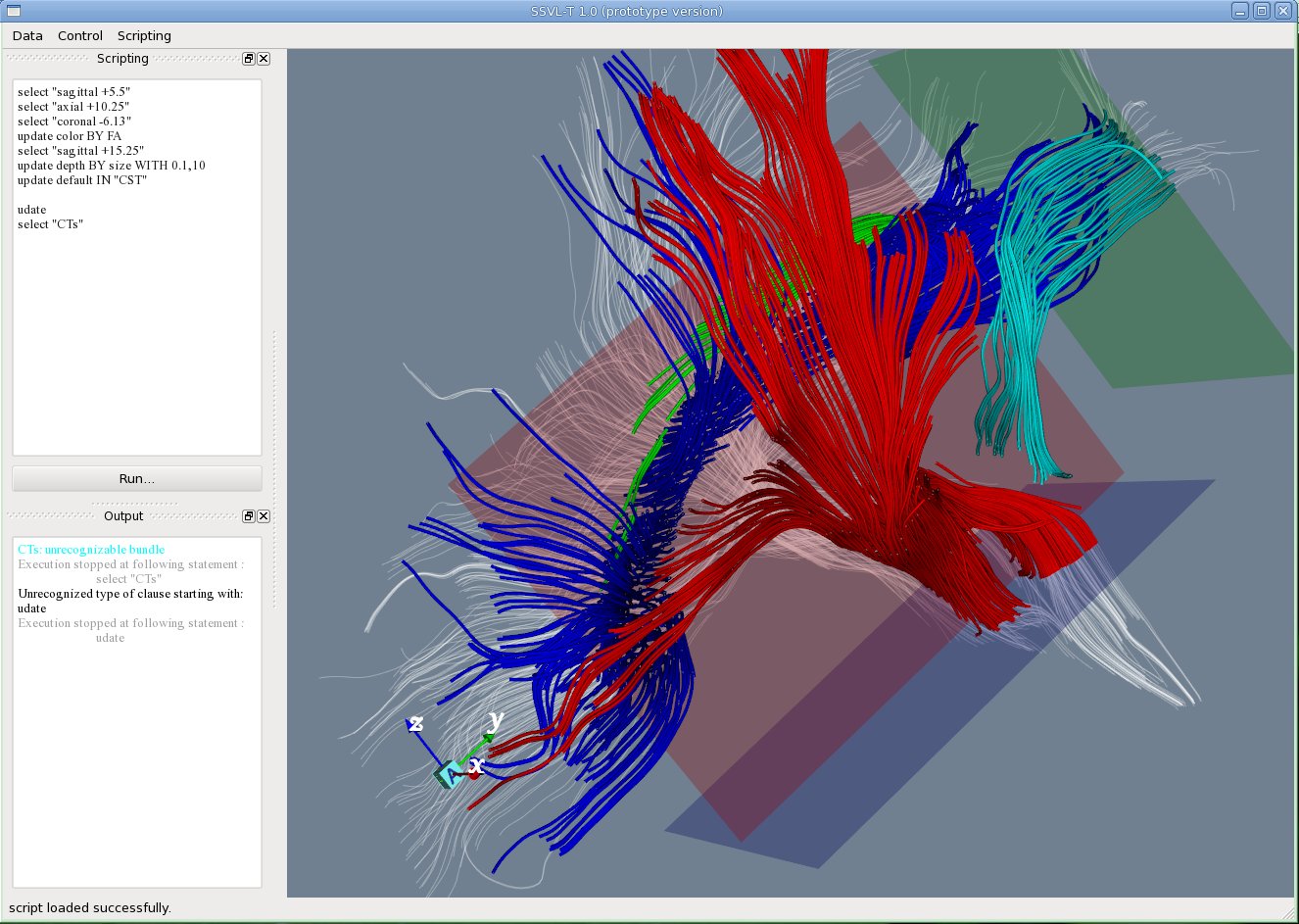}
\caption{A screenshot of the {\Tool} programming interface, consisted of a programming text board (upper left), a simple debugging
output window (bottom left) and the visualization view (right).}
\label{fig:outlook}
\end{figure}

To capture the design elements of the language, we have conducted experimental studies with domain scientists in DTI who are expected users of our language and summarized design principles for the language out of their descriptions of visualization making and exploration, from which basic lexical terms such as verbs, prepositions and conjunctions were also reduced. With these principles and language elements, we have developed a language prototype, named {\Tool}, as an initial implementation of the visualization language we are proposing. To target non-programmer users like neural medical doctors, {\Tool} is designed to be a high-level declarative language. 

Also, for an easier usage for users without any programming skills and experiences, {\Tool} is currently developed as a procedural language that contains only an intuitive type of control structure, i.e. the sequential structure. As such, users can write {\Tool} scripts simply as if they verbally describe the process of authoring visualizations in sequence. Figure~\ref{fig:outlook} gives an outlook of the {\Tool} programming interface.

The following usage scenarios briefly show the utility of {\Tool}. In the first scenario, an end user first
loads a whole DTI model and then programs to vary tube size in the default streamtube visualization
by fractional anisotropy and tube color by fiber distance to the viewing point in a specific brain structure. In the end
of his task, the user can change the streamtube representation of another brain region to ribbons.

In the second one, an user filters fibers according to an estimate of linear anisotropy threshold and then gradually
adjusts the threshold until satisfied. The user then further cuts off the selected fibers outside a target brain region through spatial commands with precisely calculated movements and thus reaches the tubes of interest.

As the final example, an user can get the size of a brain structure in terms of the number of fibers, average fractional
anisotropy in a brain region, and other common DTI metrics after reaching the target fibers.
In each of these scenarios, the user achieves each step by writing a declarative program statement in the script
editor and the results are reflected in the visualization view (see section~\ref{sec:scenarios}).

Apart from a visualization language that helps domain scientists build DTI visualizations by themselves to exactly meet their
specific needs with the visualizations, our work also contributes several design features to general DTI
visualizations including: (1) visual symbolic mapping based on color, size and shape, as is new for scientific
visualizations, (2) lexical representations of spatial relationships for 3D object visualization and
manipulations and (3) data encoding flexibility built upon the migration of Bertin's semiological principles to scientific
visualizations.

The following snippet gives a quick view of how a {\Tool} program looks like. This
script describes an exploratory process of an end user with the streamtube model ~\cite{Zang2003DTI} of a human brain DTI data set, in which different
fiber bundles (CC, CST, etc.) are filtered according to threshold of DTI metrics (LA, FA, etc.) and customized with
various visual encoding methods (shape, color, size, etc.).
\begin{Verbatim}[frame=single]
LOAD "/tmp/allfb_tagged.data"
SELECT "CC"
SELECT "FA in [0.2,0.25]" IN "IFO"
UPDATE color BY FA IN "CC"
SELECT "LA > 0.35" IN "CST"
UPDATE shape BY line IN "CC"
UPDATE shape BY tube IN "IFO"
UPDATE size BY FA WITH 0.1,20 IN "IFO"
\end{Verbatim}
As shown in this example, a {\Tool} program is essentially an intuitive sequence of steps each carrying out a single
visual transformation of data. Although the script is written in a textual form as in a traditional computer programming language, each of the statements is more like a high-level command. Also, there is no any other logic structures than the sequential one in {\Tool}, which makes this language fairly easy to learn and use for end users in medical field.

The rest of this paper is organized as follows. We first give general background and discuss related work in Section
~\ref{sec:relatedwork}. In Section ~\ref{sec:design} we detail design principles and supporting language elements
and then brief implementation issues in the following Section ~\ref{sec:implementation}. Section~\ref{sec:scenarios}
expands the details of the three usage scenarios introduced above and gives the corresponding {\Tool} scripts and running results.
We discuss other design features of our language that have not yet been fully implemented but are integral to our overall language design in Section~\ref{sec:discussion} before finally concluding the paper in Section ~\ref{sec:conclusion}.

\section{Background and Related Work}
\label{sec:relatedwork}
In this section, we describe previous work related to our visualization language design and especially compare them against {\Tool}.
\subsection{Visualization of DTI Models}
In general, DTI data set can be visualized using various approaches ranging from direct volume rendering of tensor field
~\cite{kindlmann2000strategies} to geometry rendering of the fiber model derived from tensor field. With geometry rendering,
DTI fibers are usually depicted as streamlines ~\cite{kindlmann2004visualization}, streamtubes and streamsurfaces
~\cite{Zang2003DTI}. In order to explore 3D visualizations of the fiber geometries, 2D embedding and
multiple coordinated views ~\cite{Jianu-2009-E3D} along with various interactive techniques ~\cite{Blaas-2005-FRF,Sherbondy-2005-ECB} have been employed.

Many other powerful tools have also been developed for exploring DTI visualizations
~\cite{Akers-2006-CCD,Jianu-2009-E3D,Chen-2009-ANI,Akers-2006-WOF,Blaas-2005-FRF}.
However, due to the data complexity, domain users' needs for performing their various tasks in daily practice have not
yet been fully satisfied by using those tools. To give users more flexibility, some of the visualization tools are
made highly configurable by allowing a wide range of settings ~\cite{Toussaint-2007-MedINRIA,Sherbondy-2005-ECB}.
Nevertheless, it is still challenging to design a thoroughly effective visualization tool to meet all the needs of
users. For instance, although sometimes able to meet specific requirements, higher flexibility of a visualization tool
may even make the tool more complex to use for domain users ~\cite{li2006scalable}.

\subsection{Composable Visualizations}
Since pioneered the automatic generation of graphic representation ~\cite{mackinlay1986automating}, Mackinlay's work
has been extended lately into a visual analysis system armed with a set of interface commands and defaults representing
the best practices of graphical design ~\cite{mackinlay2007show}, upon which a commercial software Tableau was developed. In his work, the generation of visualizations was automated thanks to the application of a series of design rules
and made adaptable to users with a wide range of design expertise via constrained flexibilities by those design rules.
With {\Tool}, we also intend to provide an environment in which end users can flexibly build their own visualizations like Tableau. 
However, instead of targeting visual analysis in the context of two-dimensional (2D) information visualization, {\Tool} primarily aims at end-user visualization making and exploration with 3D scientific data such as DTI. Also, compared to the visual specifications in Tableau like those in its predecessor Polaris ~\cite{Stolte-2002-Polaris}, textual programming is the main means for end users to interact with visualizations of interest in {\Tool}. Similar to Polaris in terms of using visual operations to build visualizations, the tool designed in ~\cite{Sherbondy-2005-ECB} aims to support retrieving DTI fibers instead of querying relational database in Polaris.

As a toolkit, ProtoVis gives users high-level usage flexibility even programmability yet imposes constraints upon user
programs through implicit rules to produce effective visualizations ~\cite{bostock2009protovis}. This tool has been
evolved into its descendant named D3 ~\cite{bostockd3} for a better support of animation and interaction.
{\Tool} shares some Protovis features like addressing non-programmer audience and having concise and easy-to-learn grammar. However, different from Protovis that uses simple graphical primitives called marks to construct information visualizations and mainly targets web and interaction designers, {\Tool} targets neuroscientists instead and enables them not only
to flexibly construct but effectively explore in the context of scientific visualizations exemplified by that of DTI data.

\subsection{Visualization Languages}
Processing ~\cite{fry2004computational} is more a full-blown programming language and environment than a traditional visualization tool.
Built with the full Java programming language facilities, Processing integrates the underlying visual design rules to help user build beautiful yet informative visualizations with the support of interaction design. Although developed to be accessible for new users and non-programmers, Processing is more oriented to users with certain level of programming skills and might be still challenging for domain users like neuroscientists who are the primary audience we address. A sister visual programming language of Processing, Processing.js~\cite{processing:js} also targets web developers.
By contrast, {\Tool} is distinct in that it empowers end users to explore scientific data through intuitive syntax within a sequential structure rather than offering a full set of programming features in a traditional computer language as Processing does. Like {\Tool}, Impure~\cite{impure} is also a programming language for data visualizations that targets non-programmers. Although supporting various data sources, this completely visual language is developed for information design and rather than for scientific visualizations.

Although a natural language like WordsEye~\cite{Coyne-2001-WordsEye} for visualizations might be appealing to ordinary users without any programming knowledge, we do not attempt the entirely descriptive nature for {\Tool} as WordsEye did at current stage. In terms of lexical and syntax design, {\Tool} is similar to Yahoo!'s Pig Latin~\cite{Olston-2008-PLN}, which is a new data processing language associated with Yahoo! Pig data handling environment that balances between a declarative language and a low-level procedural one. The language supports data filtering and grouping with parallelism by its map-reduce programming capability. However, this language did not handle visualizations or any form of graphical representations but focusing on ad-hoc data analysis. Also, {\Tool} sets it apart from Pig Latin in the target audience again since the latter mainly served software engineers.

The Protovis specification language~\cite{heer2010declarative} is a declarative domain-specific language (DSL) that
supports specification of interactive information visualizations with animated transitions, providing an
approach to composing custom views of data using graphical primitives called marks that are able encode data by dynamic
properties, which is similar to the mapping of object properties to graphical representations in another InfoVis
language presented by Lucas and Shieber~\cite{lucas2009simple}. To some extent, both languages are comparable to the
Microsoft's ongoing project Vedea aimming at a new visualization language~\cite{Vedea} in terms of syntactic design and
programming style, although its design goals are closer to that of Processing.

Also in the InfoVis domain, Trevil~\cite{trevil} is a programming language based on its predecessor Trevis~\cite{Adamoli10},
a framework used for context tree visualization and analysis. It supports composing visualizations but dedicates to the visualization of unordered trees. Another specific-purpose language is one presented in~\cite{peterson:fdpe02} that serves the composition of visualizations of mathematical concepts like those in basic algebra and calculus.

Recently Metoyer et. al. ~\cite{Metoyer2012UVL} report from an exploratory study a set of design implications for
the design of visualization languages and toolkits. More specifically, their findings inform visualization language design through the way end users describe visualizations and their inclination to using ambiguous and relative, instead of definite and absolute, terms that can be refined later via a feedback loop provided by the language. Emphatically, their findings also disclose that end users tend to express in generally high-level semantics. During the design of our visualization language, we have benefited from these findings and actually have reflected them in the development of {\Tool}.

\section{Language Design}
\label{sec:design}
In this section, we first summarize end-user descriptions on composing DTI visualizations from which
design requirements and principles, as follow a summary of the language symbols and
description of {\Tool} data model, are extracted and motivated respectively.
The development of {\Tool} is driven by end-user
requirements with DTI visualizations and the design principles are embodied in the language features of {\Tool}.
After each of the language features, {\Tool} language elements that meet the feature are detailed, including related
lexical terms and syntactic patterns.
Instead of describing the implementation techniques, which are briefed in Section~\ref{sec:implementation}, this section
emphasizes how the design principles and language elements address the end-user requirements.

\subsection{Design Motivations}
\label{sec:motivation}
The design of {\Tool} is motivated by the needs of typical end users we target for composing DTI
visualizations by themselves, which can be derived from their verbal descriptions about visualizations they would
desire in our many interviews and discussions with them. We report just a few representative example comments of them
on visualizations produced beforehand by computer scientists.

Our participants include neurologists and neural physicians, both conducting clinical diagnosis with DTI data
visualizations. In a typical interview, participants are presented visualizations of a same DTI brain data set composed
differently by manipulating various visual elements and the compositions are done by computer scientists, who then
revise the composing process according to the comments of participants. As results, either the unsatisfactory visualizations
are finally modified to meet participants' requirements or suggestions for achieving the desirable visualizations are
received if current tool is not capable of composing the desirable ones.

As an example, multiple visual mappings of depth values to size and color does not enhance
the visualization of DTI model as expected. Surprisingly, ``\textit{...it is misleading to have the different size}" while
color has already been used to discern depth, and ``\textit{...would rather have it stay the same size as I spin it
around.}". However, visual mapping of depth to color is still preferable since ``\textit{...I like it with the color.
That is what I need to look at}". Nevertheless, the composed coloring scheme in which color is mapped by depth might be
also useful ``\textit{...if determined by the principal eigen values}". And, ``\textit{...I think that color is a good
idea but prefer color by orientation...}", etc.

There is also a call for doing analysis in the composing environment (``\textit{...Also, one thing for fibers, I am looking at for analysis purposes}"). Emphatically, both classes of participant unanimously ``\textit{want to do the analysis over here on the same page, that will be good, too, rather than opening it up again and trying to do it... It will all come together. It will all be integrated into one...}".

These observations all suggest that domain users, exemplified by the typical end users of {\Tool} above, potentially ask
for a high-level tool allowing them to define a self-control sequence of operations that works towards a visualization
precisely meeting their own specific needs. By allowing users to compose with well-designed visual elements,
a programming environment can provide the capabilities for neurologists to create their own visualizations,
by which our present work is justified.

Furthermore, our work with {\Tool} is substantially grounded upon the semiology of the graphic
sign-system and especially the taxonomy about the properties and characteristics of retinal variables ~\cite{bertin1983semiology}
in terms of the syntax and semantics design for the scientific visualization language. {\Tool} incorporates a subset of
the properties and characteristics that are most relevant, according to neurologists' verbal descriptions about DTI
visualizations, to the language structure and content: size variation, color variation and shape variation. For one
thing, corresponding syntax terms are built into the language core as basic symbols. For another, semantics associated
with these terms are designed to support composing DTI visualizations with respect to these retinal variables by
allowing free manipulations of the attributes of related variables. 

While the semiology and taxonomy is originally
formulated to guide the design of 2D graphical representations, we extend them into the 3D graphical environment
and employ in the case of DTI visualizations. Further, we expand the scope of this taxonomy particularly for 3D
visualizations by including a dimension related to depth perception, called "depth separation" in our language design
in addition to the legacy retinal dimension. Correspondingly, composing the depth separation is enabled through built-in
support in {\Tool}. Primary visual elements such as value, transparency, color and size are employed once again but now
for the purpose of the depth-dimension composition.

It is fairly noteworthy, and common as well, in participants' verbal descriptions that spatial terms are frequently used
and most of the terms related to spatial locations are relative rather than measured in precise units. That
{\Tool} is designed to be a spatial language is exactly in response to the concerns of our target end users with the
spatial relationship of data components in the scientific data model being visualized. The participants' descriptions
are also in accordance with the fact that spatial constraint is a defining data characteristic scientific visualization.
Consequently, {\Tool} includes a set of syntactic and semantic supports for spatial operations in order to meet end-user
needs for composing 3D scientific visualizations like that of DTI models.

Intending to be an initiative of an end-user programming approach to scientific visualizations, {\Tool} is designed to support an environment in which domain scientists as end users can compose highly customizable visualizations reflecting their thinking process with the graphical representations of their data set. Since {\Tool} is incubated from DTI
visualizations, the language design primarily deals with DTI data. In this context, language elements of the present
{\Tool} are derived from experimental study with neuroscientists using diffusion MRI data models. As a matter of fact,
the symbols and syntax of current {\Tool} are extracted from verbal descriptions of neurologists using DTI
about how they would create and explore DTI visualizations. As we often refer as end users, neuroscientists,
neurologists and other medical experts who conduct clinical practice with DTI data and its visualizations are the primary audience our {\Tool} language targets.

\subsection{Language Symbols}
The core content of {\Tool} itself is a simple set of language symbols and keywords. End-user actions intended with DTI
visualizations are triggered through five key verbs that are all complete words in natural English. Prepositions are
used for targeting scope of data of interest and conjunctives for connecting statement terms. All operators used in {\Tool} are exactly the same as those used in elementary math. Specifically, $[]$ serves range operator here for giving a numerical bound that is used in conditional expressions and $+$ and $-$ are relative (increment and decrement) operators rather than serving arithmetical operations (addition and subtraction). Several built-in routines are provided in {\Tool} for simple data
statistics and analysis in DTI visualizations: $AvgFA$ and $AvgLA$ calculates the average FA and average
LA of a scope of fibers respectively, and $NumFiber$ stats the number of fibers in a fiber bundle.
Among the reversed {\Tool} constants, the aforementioned five major fiber bundles in human brain model are included.

In these language symbols, all verbs and prepositions are directly picked up from our neurologist collaborators' common
descriptions of visualization composition and exploration in natural language. Fiber bundle constants are also suggested
by them and operators, built-in routines and other constants are reduced from our requirement analysis of their verbal descriptions. As shown in the Table~\ref{tab:symbols}, current {\Tool} implementation contains a small set of symbols. However, our language has been designed to be scalable to increase in each type of the symbols listed in terms of implementation techniques.
\begin{table}[ht]
	\caption{{\Tool} language symbols and keywords}	
    \begin{tabular}{|p{3cm}|p{4.5cm}|}
	\hline
	Verbs & LOAD, SELECT, LOCATE, \newline UPDATE, CALCULATE \\ \hline
	Prepositions & IN, OUT \\ \hline
	Conjunctives & BY, WITH \\ \hline
	Operators & [], $<$, $<$$=$, $>$, $>$$=$, $==$, $=$, $+$, $-$ \\ \hline
	built-in routines & AvgFA, AvgLA, NumFiber \\ \hline
	Constants & shape, color, size, depth, FA, LA, \newline sagittal, axial, coronal, \newline CC, CST, CG, IFO, ILF \newline DEFAULT, RESET \\
    \hline
	\end{tabular}
	\label{tab:symbols}
\end{table}

\subsection{Data Model and Input}
To meet the design goals when targeting our domain end users, our visual language is intended to be straightforward for programming. Therefore, {\Tool} does not involve any distinction of specific data types. It does not require uses to deal with any low-level data processing procedures either. Instead, {\Tool} focuses on visual transformations in 3D visualization. As previous examples disclose, we have used a classified geometrical data model derived from DTI volumes, in which fibers are clustered in terms of brain anatomy. In our present data model as input to {\Tool}, each fiber has been manually tagged with anatomical cluster
identity as one of the five major bundles. In practice, {\Tool}'s ability to recognize the constants for the major
anatomical bundles depends on these cluster tags in the structure of the data model input. However, our language
design is not restricted to only handling clustered data. Actually, {\Tool} is freely adaptable to an unclustered data
model, although data target specification with the major bundle constants will be processed as the whole model then.
Nevertheless, {\Tool}'s capability of spatial operations empowers users to explore ROIs in the unclustered data models.

In a {\Tool} program, the first step is to indicate the source of data model by giving the name of a data file.
As an example, a {\Tool} data input statement is written as:
\begin{Verbatim}
normalBrain = LOAD "data/normalS1.dat"
\end{Verbatim}
where the LOAD command parses the input file and creates data structures that fully describe the data model, including to
identify the cluster tags. This input specification statement can also update current data model at the beginning of
the visualization pipeline if it is not the first step in a {\Tool} script. The evaluation is optional and, when provided,
saves the result to a variable ($normalBrain$ here) for later references. This is not used in current version of {\Tool}
but is required for exploring multiple data sets concurrently (see Section~\ref{sec:discussion}).

\subsection{Task-driven Language}
The language design of {\Tool} is originally driven by the visualization tasks that domain users need to perform in
their ordinary clinical practices. Among others, some of their typical tasks are (1) checking integrity of neural
structures of a brain as a whole, (2) examining fiber orientation in a region of interest (ROI) or fiber connectivity across ROIs, (3) comparing fiber bundle sizes between brain regions, (4) tracing the variation of DTI quantities such as FA
along a group of fibers and (5) picking particular fibers according to a quantitative threshold, etc.

When using DTI visualizations, not only looking for the whole data model, neurologists are also inclined to concentrate on regional details. In the case of brain DTI visualizations, they often narrow down the view scope toward a relatively large anatomical area in the first place and then dive into a specific ROI. In other words, they tend to pay more attentions to ROIs than to the whole brain.
More specifically, in the visualizations where neural pathways are depicted as streamtubes, the ROIs are usually clusters
of fiber tracts called fiber bundles. For instance, at the beginning of a visualization exploration, one of our neurologist collaborators intends to look into frontal lobe fibers within the intersection of two fiber bundles, CST and CC, and ignores all other regions of the model.
Further, suspicious of fibers with average FA under 0.5 for a cerebral disease with which the brain is probably afflicted, the user goes on to examine exactly the suspect fibers. Later on, the user focuses on the small fiber region to see how it differs from typical ones, in terms of orientation and DTI metrics, say.

\begin{figure}[htb]
\centering
\includegraphics[width=8cm,height=7cm]{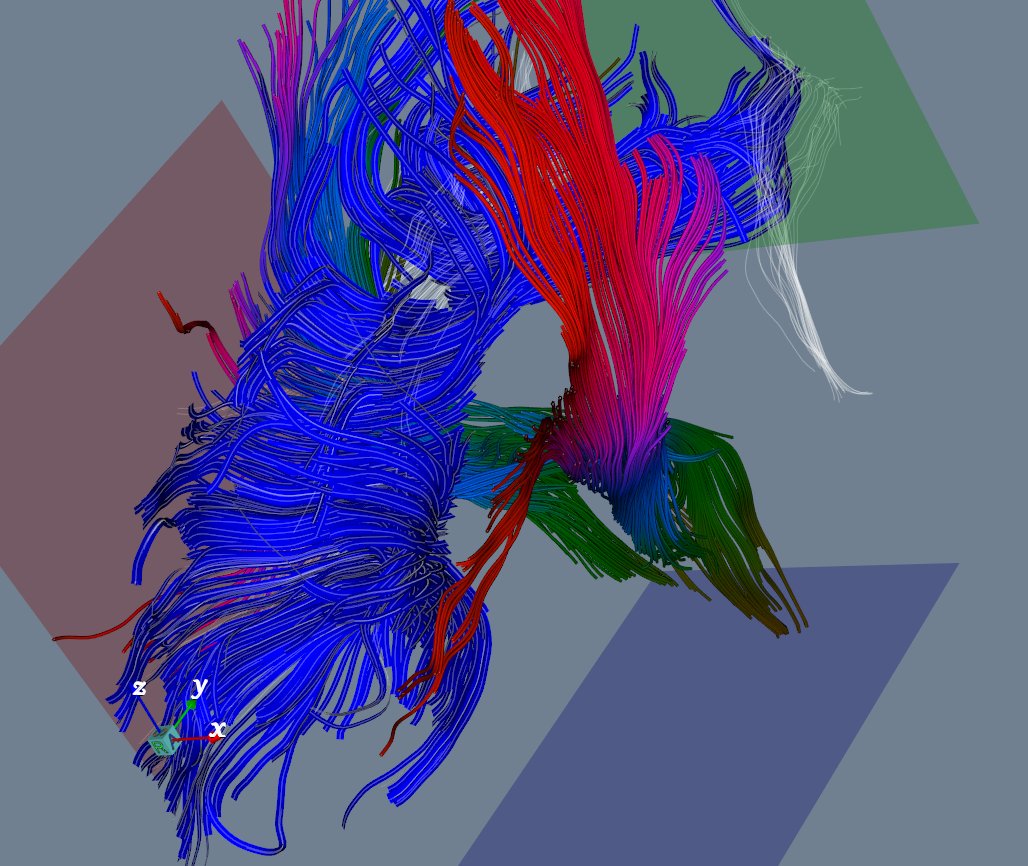}
\caption{Illustration of the task-driven design of {\Tool}.}
\label{fig:taskdriven}
\end{figure}

{\Tool} is designed as a task-driven language to support this requirement process through high-level primitives such as
SELECT and common arithmetical conditional operators including a range operator $in$. {\Tool} is mainly featured with
facilities for step-by-step data filtering with these primitives. For example, suppose the user above is to explore the fibers of interest, he can write in {\Tool} as:
\begin{Verbatim}[frame=single]
SELECT "FA < 0.5" IN "CST"
SELECT "FA < 0.4" IN "CC"
\end{Verbatim}
As the result, fibers in both interested bundles with average FA under 0.5 will be highlighted to help users focus on the
local data being explored. On top of this, the user can customize the visualization of the filtered fibers through various visual encoding methods using the UPDATE syntax. This is particularly useful when he wants to keep the data already reached in focus before moving to explore other relevant local data in order to add more fibers into his focus area, or when he simply seeks for a more legible visualization of the data firstly reached. The instance below, following the same example, illustrates how a better depth perception achieved by a type of depth encoding, together with a differentiating shape encoding, are added up to the two selected fiber bundles respectively.
\begin{Verbatim}[frame=single]
SELECT "FA < 0.5" IN "CST"
SELECT "FA < 0.4" IN "CC"
UPDATE depth BY color IN "CST"
UPDATE shape BY ribbon IN "CC"
\end{Verbatim}
This simple sequence of commands help users locate desirable fiber tracts with high accuracy while allowing flexible
customization upon current visualizations. With this language, users compose intuitive steps to finish their tasks that are
difficult to achieve by visual interactions. In this case, tracts of interest (TOIs) are first focused and then further
differentiated for more effective exploration through improved legibility. In general, {\Tool} design emphasizes this
task-driven process of visualization exploration, which fits the thinking process of end users with the present visualizations.
Figure~\ref{fig:taskdriven} shows the resulting visualization.

Filtering data in order to reach a ROI is an operation quite, nearly the most, frequently used during our neurologist
collaborators' explorations in DTI visualizations. {\Tool} offers two commands for data filtering: SELECT and LOCATE. The data filtering syntax
pattern in {\Tool} is:
\begin{Verbatim}[frame=single]
SELECT condition|spatialOperation
        IN|OUT target
result = LOCATE condition IN|OUT target
\end{Verbatim}
With similar functionality, these two commands have different semantics: SELECT executes filtering in an immediate mode by
highlighting target fibers while LOCATE commits an offline filtering operation, retrieving target fibers and sending the
result to a variable without causing any change in the present visualization. Also, SELECT provides relative spatial operations through moving anatomical cutting planes. In fact, it is tempting to combine these two commands into one while
differentiating the two semantics (by recognizing the presence of variable evaluation and taking spatial operations as
an alternative to the $condition$ term). However, we still keep these two commands separately based on end-user comments
asking for a more straightforward understanding of the semantics and easier memory of language usage. For example,
\begin{Verbatim}[frame=single]
SELECT "LA <= 0.72" IN "ALL"
partialILF = LOCATE "FA in [0.5,0.55]"
            OUT "ILF"
\end{Verbatim}
The SELECT statement will filter fibers in the whole DTI model with average anisotropy greater than 0.72 (by putting
them in the contextual background) and highlight all other fibers. In comparison, the LOCATE statement will not update
the visualization but pick up fibers outside the ILF bundle having average FA value in the specified range. Note that
when no specific data encoding applied, different colors will be applied to ROI fibers in different major bundles in
{\Tool} for discerning one ROI from another when there is more than one highlighted. Also, filtered fibers will still be
in semi-transparency as the contextual background rather than being removed from the visualization.

\subsection{Data Encoding Flexibility}
According to Bertin's semiotic taxonomy ~\cite{bertin1983semiology}, graphically encoding data with
key visual elements such as color, size and shape play a critical role in the legibility of 2D graphical representations.
In 3D visualizations, occlusion effect, an import factor considered in depth perception, has detrimental effects on impact on the overall legibility, and depth cue (DC) is an ordinal dimension in the design space of 3D occlusion management for the visualizations~\cite{elmqvist2008taxonomy}.

Therefore, we combined both aspects in our {\Tool} language design: symbolic mapping of color, size and shape for 2D graphical legibility enhancement and depth encoding, also via common visual elements such as color, size, value (amount of ink) and transparency, as depth cues for occlusion reduction in the 3D environment. As already shown in the previous example
scripts, {\Tool} allows end users to freely customize DTI visualizations using either a single data encoding scheme alone or
compound encoding scheme by flexibly combining multiple encoding methods. The latter leads to a mixed visualization as
illustrated in Figure~\ref{fig:outlook}.

\begin{figure}[htb]
\centering
\includegraphics[width=8cm,height=7cm]{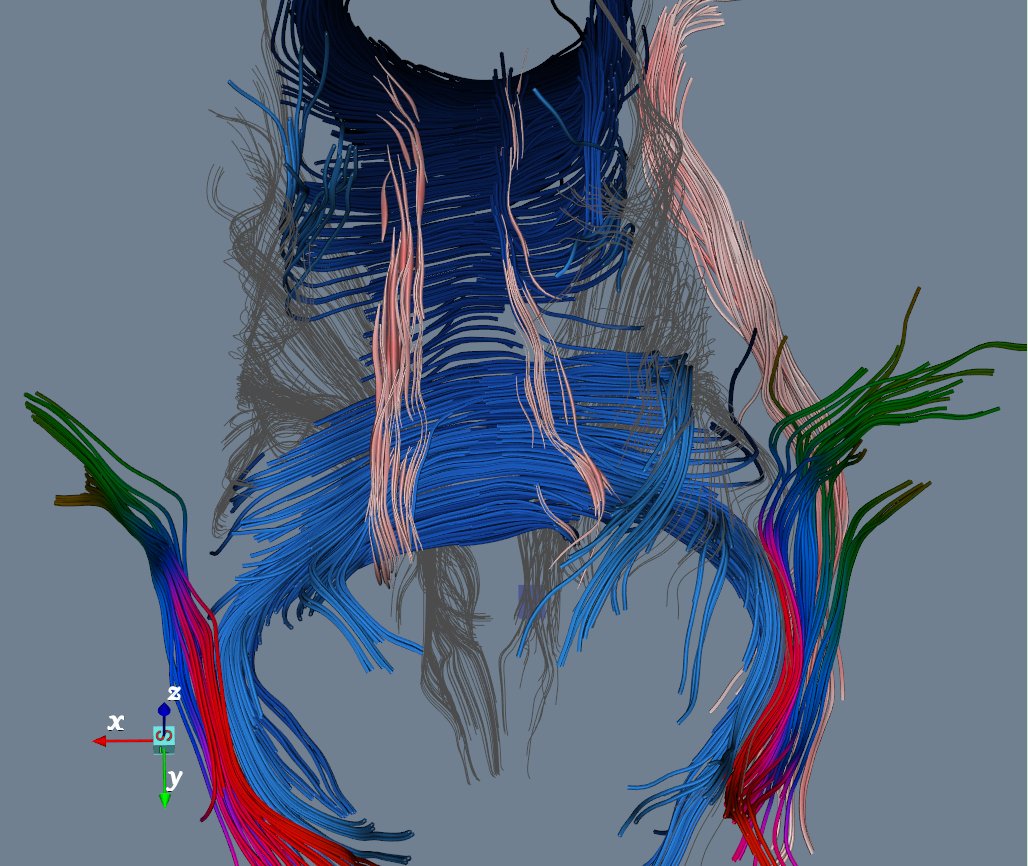}
\caption{Flexible data encoding built in the design of {\Tool}.}
\label{fig:dataencoding}
\end{figure}

In their composing or exploratory process with DTI visualizations, users often attempt to examine more than one data focus
simultaneously and would like to differentiate one focused ROI from others so that they will not get lost themselves within
the multiple ROIs. There are also other occasions under which the users have difficulty in navigate along the depth dimension
even in a single ROI. The data encoding flexibility in {\Tool} is driven by both of the two user attempts. For an
example, suppose a user has composed the streamtube visualization of a brain DTI data set with default data encoding (uniform
size, color and shape without depth cues) and now wants the overall encoding scheme to be different across fiber bundles. In order to achieve this effect, an example {\Tool} snippet can be written as follows:
\begin{Verbatim}[frame=single]
SELECT "ALL"
UPDATE shape BY LINE IN "CST"
UPDATE size BY FA IN "CG"
UPDATE color BY FA IN "IFO"
UPDATE depth BY transparency IN "CG"
UPDATE depth BY value IN "CC"
       WITH 0.2,0.8
UPDATE depth BY color IN "ILF"
\end{Verbatim}
Then, in the resulting visualization, each of the five major bundles will be visually disparate from others since all
these bundles are encoded differently.
Figure~\ref{fig:dataencoding} shows the resulting visualization.

Oftentimes, once one ROI or more filtered out, it is also necessary to examine the selected fibers more carefully. For this
purpose, {\Tool} allows users to impose various data encoding schemes upon data targets. Such visualization customization is
done by the UPDATE command, which always works in an immediate mode causing update in the current visualization after
execution. The general UPDATE syntax pattern is:
\begin{Verbatim}
UPDATE var1 BY var2
  WITH para1,...,paraN IN|OUT target
\end{Verbatim}
where $var1$ indicates an attribute, such as shape, color, size, depth, etc., of current visualization to be modified, and $var2$ gives how the actual updating operation is to be performed in terms of its relation to $var1$. The parameter list ending the statement presents extra information the updating requires, as is specific to a particular data encoding operation. Like the target specification (optional with all commands as stated before), the BY clause and WITH clause are both optional. Table ~\ref{tab:updatesyntax} lists all possible combinations of $var1$, $var2$ and associated parameter list already developed in present {\Tool}.
\begin{table}[ht]
	\caption{Combination rule of constants in UPDATE statement of current {\Tool} implementation}
	\begin{tabular}{|p{2cm}|p{3cm}|p{2cm}|}
	\hline
	$var1$ & $var2$ & $parameters$ \\
    \hline
	\hline
	shape & line, tube, ribbon & N/A \\ \hline
	color & FA, LA & N/A \\ \hline
	size  & FA, LA & minimal,scale \\ \hline
	depth & size,color,value,\newline transparency & lower,upper \\ \hline
	DEFAULT & N/A & N/A \\ \hline
	RESET	& N/A & N/A \\
    \hline
	\end{tabular}
	\label{tab:updatesyntax}
\end{table}
In the table, ``lower,upper" gives the bound of depth mapping and ``minimal,scale" indicates the minimum and
the scale of variation in size encoding. DEFAULT and RESET, when going with the verb UPDATE, act as a command for revoking all data filtering and data encoding operations respectively.
The following script shows how to inspect the change of FA along fibers in a ROI by mapping FA value to tube size, which
results in a more intuitive perception of the FA variation in that ROI.
\begin{Verbatim}[frame=single]
UPDATE RESET
partialILF = LOCATE "FA in [0.5,0.55]"
            OUT "ILF"
UPDATE size BY FA IN "partialILF"
\end{Verbatim}

\subsection{Spatial Exploration}
One of our main design goals with {\Tool} is to provide a language with which users are able to operate spatial structures.
We found that our neurologist collaborators tend to frequently use spatial terms such as ``para-sagittal", ``in", ``out", ``mid-axial" and ``near coronal", etc. in their descriptions about DTI visualizations in the 3D space. They also use a set of other general spatial terms including ``above", ``under", ``on top of", ``across" and ``between", etc. like those found in ~\cite{Metoyer2012UVL} and more domain-specific ones such as ``frontal", ``posterior" and ``dorsal", etc. At present stage, {\Tool} only contains a subset rather than all of these spatial terms.

In such a 3D data model as that from DTI, spatial relationships between data components are one of the
essential characters, which are actually typical of 3D scientific data in general.
Accordingly, composing a DTI visualization necessitates the capability of using spatial operators with domain
conventional terms in order to describe the process of visualization authoring. In response, {\Tool} supports spatial
operations through two approaches combined. First of all, three visible cutting planes that help guide in the three conventional anatomical views, namely the axial, coronal and sagittal view respectively, are integrated in the visualization view (see Figure ~\ref{fig:outlook}). Then, flexible manipulating operations upon the three planes are built into the {\Tool} spatial syntax definitions. This enables end users to navigate in the dense 3D data model with a highly precise filtering
capability exactly as they examine a brain model in clinical practice.

\begin{figure}[htb]
\centering
\includegraphics[width=8cm,height=7cm]{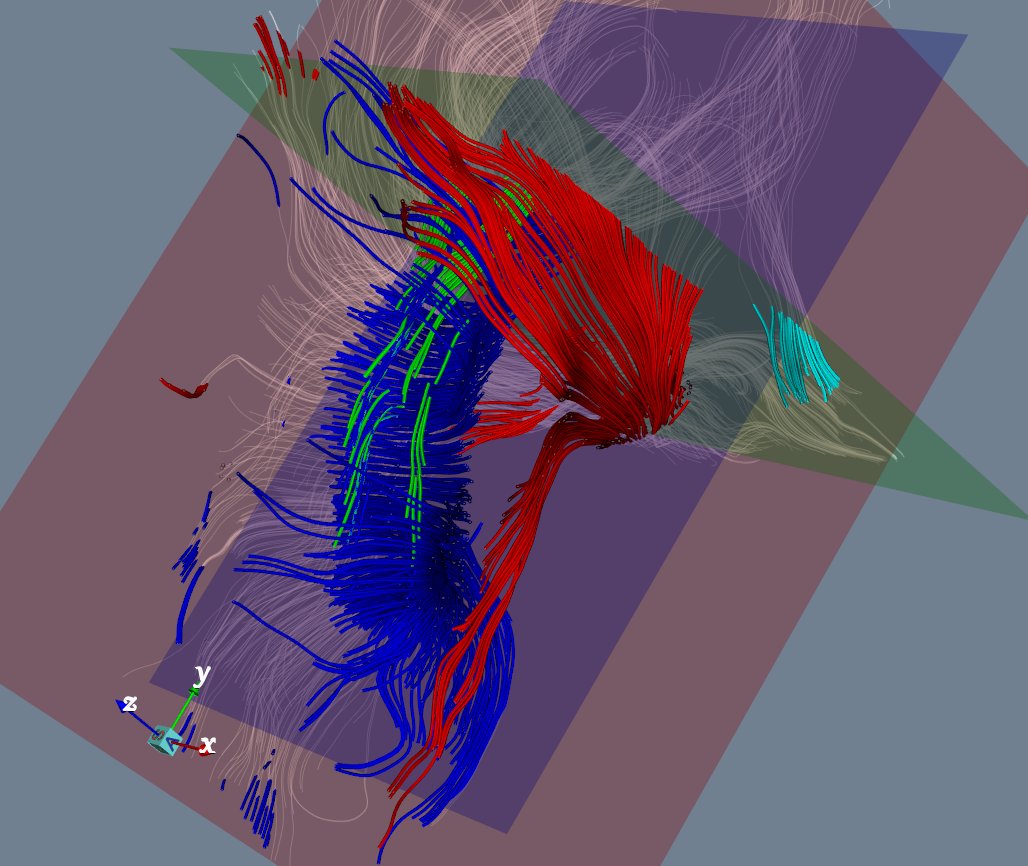}
\caption{Illustration of the design of {\Tool} as a spatial language.}
\label{fig:spatial}
\end{figure}

For instance, suppose the streamtube representation of a DTI model being programmed is derived using unit seeding
resolution from DTI volumes with a size of $\displaystyle 256 \times 256 \times 31$ captured at a voxel resolution of
$\displaystyle 0.9375mm \times 0.9375mm \times 4.52mm$, and
suppose both the axial and coronal planes are located at their initial position so that nothing is cut along these two
views. In order to examine suspect anomaly in the brain region of occipital lobe, a medical doctor attempts to filter the
data model as such that approximately only this region will be kept. For this task, the corresponding {\Tool} script can be written as:
\begin{Verbatim}[frame=single]
SELECT "coronal +159.25"
SELECT "axial -27.5"
\end{Verbatim}
Similarly, relative movements can be imposed on the sagittal plane as well. These simple relative operators included in
{\Tool} in support of spatial exploration is also informed by the design implications given in ~\cite{Metoyer2012UVL}
although mainly comes from user requirements of performing DTI visualization tasks pertaining to spatial operations.
Figure~\ref{fig:spatial} shows the resulting visualization.

\subsection{Flat Control Structure}
Our another main design goal with {\Tool} is to provide a declarative language environment for domain end users who have neither any programming skill and experience nor basic understanding of computer program structures. Consequently, we purposely eliminate the conditional and iteration structures from the language design of {\Tool} and only keep the most intuitive one, i.e. the sequential structure, since this simple structure is much more intuitive than the other two. This features {\Tool} with a flat control structure that is essential for achieving the design goal. Alternatively, {\Tool} uses high-level semantics to overcome its otherwise weakness in expressing user task requirements for lack of these two missed control structures through two approaches addressing the requirements for them.

First, requirement for an iteration structure usually stems from the needs to operate on multiple targets.
Here in {\Tool}, operation target is a common term in all syntax patterns to indicate the scope of data to focus on. We
address this requirement through enumeration and target term defaults in {\Tool} syntax patterns. On the one hand,
with enumeration, end users simply list all targets in the target term to avoid iteration. For example, suppose a user intends to select three bundles and then to change size encoding for two of them, his {\Tool} script can include:
\begin{Verbatim}[frame=single]
SELECT "CST,CC,CG"
UPDATE size BY FA IN "CST,CG"
\end{Verbatim}
As such, no iteration structure for looping through the multiple targets is needed.
On the other hand, with term default, when missing a target term in single statement, "ALL" will be
assumed as a default scope meaning the whole data model to be the target. This rule is applicable for all types of {\Tool}
statement, which means that target term is optional in all {\Tool} syntax patterns.

\begin{figure}[htb]
\centering
\includegraphics[width=8cm,height=7cm]{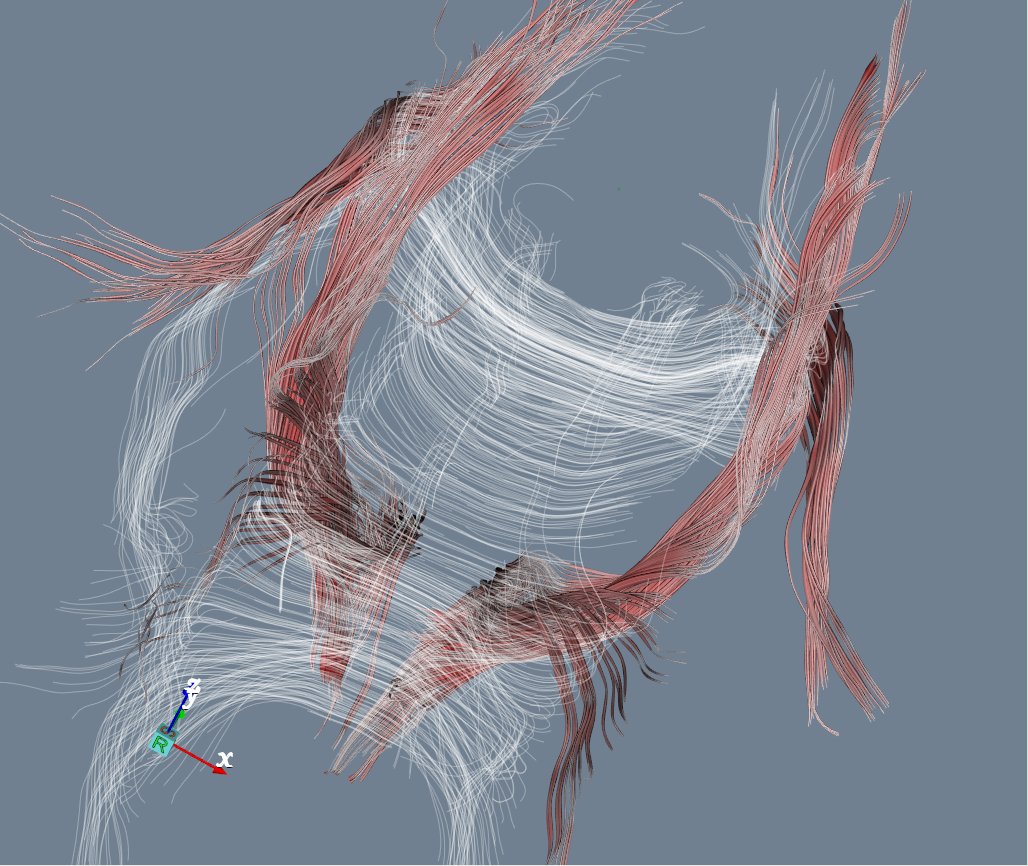}
\caption{Illustration of the flat control structure of {\Tool} program.}
\label{fig:flat}
\end{figure}

Second, requirement for a conditional structure comes from users' requests for a means to express conditional processing.
For example, they often filter fibers according to FA thresholds. In {\Tool}, conditional expression can be flexibly embedded in a statement to avoid this structure. It has been shown in previous examples how to embed conditional expressions in SELECT statement. For syntactic simplicity, condition is expressed in UPDATE statement indirectly through variable reference as the following another example snippet shows.
\begin{Verbatim}[frame=single]
suspfibers = LOCATE "FA in [0.2,0.25]"
             IN "CST,ILF"
UPDATE size BY FA IN "suspfibers"
\end{Verbatim}
where LOCATE is an alternative to SELECT but it results in a storage of the fibers filtered into a variable for later
reference instead of highlighting those fibers immediately as SELECT does (see Section~\ref{sec:design} for detailed language
elements).
Figure~\ref{fig:flat} shows the resulting visualization.

\subsection{Fully Declarative Language}
\begin{figure}[htb]
\centering
\includegraphics[width=8cm,height=7cm]{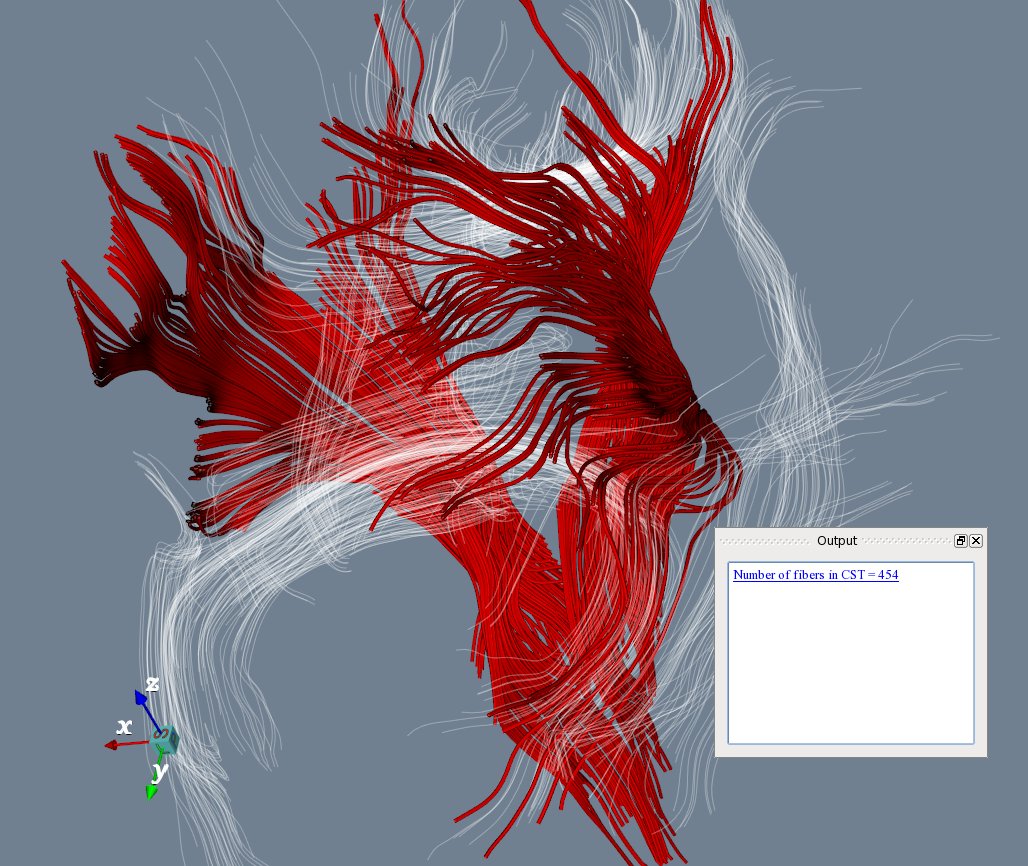}
\caption{Result of an example script showing {\Tool} as a fully declarative language.}
\label{fig:declarative}
\end{figure}

Since the end users of {\Tool} are medical experts who prefer natural descriptions to programming style of thinking
according to our talks to them, elements even merely close to those in a computer programming language
have been changed to be as declarative as possible. In {\Tool}, all types of statement are designed to be in a consistent
pattern: started with a verb, followed by operations and ended by, optionally, data target specification, with optional
evaluation of statement result to a variable for later reference if provided. This syntax consistency has been applied to
even the data measurement statement where invocation of built-in numerical routines is involved. To measure the number
of fibers in a selected bundle, for instance, instead of writing as:
\begin{Verbatim}
CALCULATE  NumFibers("CST")
\end{Verbatim}
users with {\Tool} write
\begin{Verbatim}
CALCULATE  NumFibers IN "CST"
\end{Verbatim}
In addition, all keywords in {\Tool} are case insensitive in order to reduce typing errors. Neuroscientists comment that
these features make the language easy to learn and intuitive to use.
Figure~\ref{fig:declarative} shows the resulting visualization.

As exemplified above, besides visually examining the graphical representations, medical experts often need to investigate the DTI data itself
in a quantitative manner. In clinical practice of neuroscientists using DTI, quantities such as
average FA and number of fibers are important DTI tractography-based metrics for assessing cerebral white matter
integrity ~\cite{correia2008quantitative}. In fact, these metrics are usually used in our end-user description of DTI visualizations as well.
Accordingly, {\Tool} provides capabilities to calculate some DTI metrics most frequently used in end users' practice of
diagnosis through built-in numerical routines. The following pattern shows the {\Tool} data analysis syntax.
\begin{Verbatim}
val = CALCULATE metricRoutine IN|OUT target
\end{Verbatim}
At current stage of {\Tool} development, $metricRoutine$ can be one of $AvgFA$, $AvgLA$ and $NumFibers$, whose functions
have been described before. More routines can be extended later based on further comments of our end users.
In this syntax pattern, keeping the resulting value by evaluation is optional and sometimes useful when being referred to
afterwards (see usage scenario 3 described in Section~\ref{sec:scenarios}).
For example, in order to sum up fibers with average FA falling within a particular range and then figure out average LA
of the target fibers, an end user can write following script in {\Tool}:
\begin{Verbatim}[frame=single]
frontalmix = LOCATE "FA >= 0.35"
    IN "CST,CC"
CALCULATE NumFibers IN "frontalmix"
CALCULATE AvgLA IN "frontalmix"
\end{Verbatim}
After running, the script above will dump result in the output window in the {\Tool} programming environment as shown in
Figure~\ref{fig:outlook}.

\section{Implementation}
\label{sec:implementation}
{\Tool} is declarative in its general form with support of certain programming language features such as
variable referencing and arithmetical and logical operations. At this early stage, the language scripts are not
executed via a fully-featured interpreter or compiler but a string-parsing based translator of descriptive text to
visualization pipeline components and manipulations upon them. The core of {\Tool} is implemented on top of
the Visualization Toolkit (VTK) using C++. The rendering engine is driven by the visualization pipeline and
legacy VTK components ranging from various geometry filters to data mappers. However,
in order to support language features such as mixed data encoding and depth mapping in {\Tool}, a group of new pipeline
components like those for view-dependent per-vertex depth value ordering has been extended on top of related VTK
classes, and many legacy VTK components have been tailored for specific needs of visualizations in {\Tool}.

In particular, the core of {\Tool}, the script interpreter has also been implemented primarily as data filters in the VTK
visualization pipeline. For instance, filtering according to thresholds of DTI metrics is developed as a set of separate VTK
filters each serving a specific metric. As such, interpreting a {\Tool} script is to translate the text, according to
defined syntax and semantics, to data transformations in the VTK pipeline. For achieving the data encoding flexibility,
multiple VTK data transformation pipelines have been employed in current {\Tool} implementation.

Additionally, the overall programming interface is implemented using Qt for C++. For example, interactions like
triggering execution of a {\Tool} program, serializing and deserializing the text script, etc. are all developed with Qt
widgets, although the interactions with the visualization itself are handled using legacy VTK facilities with necessary
extensions. Figure~\ref{fig:outlook} illustrates the outlook of current {\Tool} programming interface. Both the code
editor and "debugging" information window are of dockable widgets, which facilitate the script programming by allowing
free positioning and resizing as opposed to the visualization view.

Since our language is definitely non-programmer oriented, program debugging skills are not expected of users.
Consequently, instead of building a full-blown debugging environment as seen in almost all integrated development
environments (IDEs), we simply use a dockable output window to prompt users all error messages caused by invalid syntax or
unrecognized language symbols. We have made use of GUI utilities of Qt for C++ to dump, after
running a script, those messages to tell what and where is wrong in natural language description with different levels of
errors (fatal, warning and notice, etc.) differentiated by different combinations of font size, type and color of the text. Resulting values out of running data analyzing statements are also displayed in this output window. We do not set a separate window for displaying those numerical results in order to simplify the programming interface and, alternatively, we use remarkably disparate text background and underscore to highlight them among other messages. Also, natural language description has been used to present those numerical results so that they are easy to read and understand for end users.

Although there is no special requirement regarding hardware platform configuration, a high-speed graphics card like those having, for instance, 512M VRAM and 50MHz GPU is preferred for rendering the dense 3D DTI data model efficiently, which makes the {\Tool} programming environment work smoothly as a whole.

\section{Usage Scenarios}
\label{sec:scenarios}
In this section, we describe several sample tasks done by neurologists with visualizations of a brain DTI model using
the {\Tool} language. The usage scenarios associated with the sample tasks are representative of some typical real-world
visualization tasks of neuroscientists and neurological physicians with expertise in DTI in their clinical practices.
The usages range from visualization customization and exploration to DTI data analysis, covering the main language features and functionalities of our current {\Tool} implementation.

In the following scenarios, Josh, an end user of {\Tool}, has a geometrical model derived from a brain DT-MRI data set
wants to compose and explore visualizations of the data for diagnosis purpose. For each of the scenarios, Josh fulfills his
task by programming a {\Tool} script that describes his thinking process for that task and then clicks the "Run" button to
execute the script. Josh programs with {\Tool} syntax references showing on a help window and corrects any term that is
typed incorrectly with the assistance of error messages displayed in the output window. Once the script is interpreted
correctly, either the visualization gets changed or numerical values coming out in the output window, as the
results of script execution. Scripts and running results are presented at the end of the description of each usage
scenario.

\subsection{Scenario 1: composing visualizations}
To start with, Josh specifies a data file that contains the geometries of the brain DTI model using the LOAD command.
As used in examples throughout this paper, the model contains five major fiber bundles that have been marked in its
storage structure in a text file: corpus callosum (CC), corticospinal tracts (CST), cingulum (CG), inferior longitudinal fasciculus (ILF) and inferior frontal occipital fasciculus (IFO). By default, running this single statement gives a streamtube visualization of the model with uniform visual encoding across all major bundles and without depth encoding. 

\begin{figure}[htb]
\centering
\includegraphics[width=8cm,height=7cm]{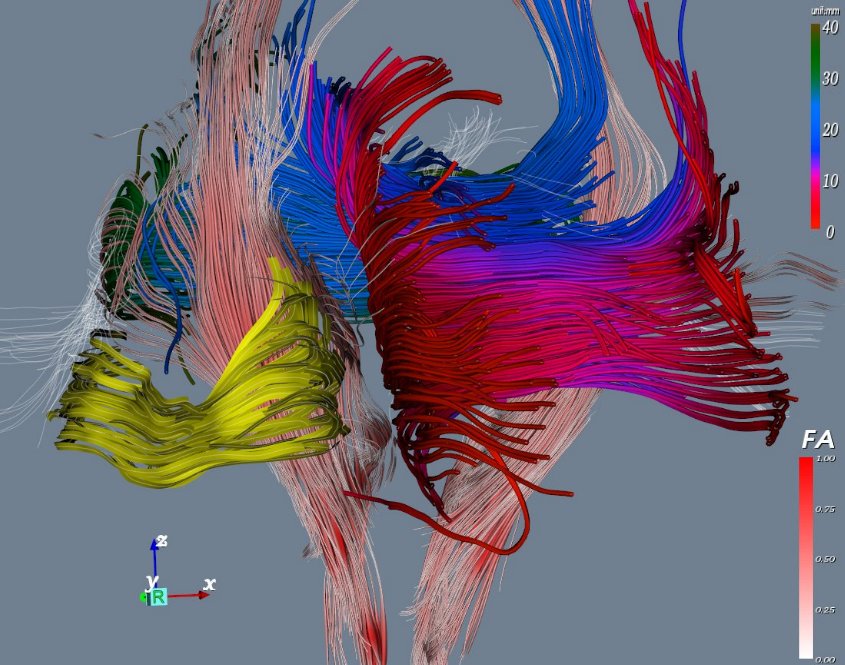}
\caption{Screenshot of the visualization resulted from running the {\Tool} program written in scenario 1.}
\vspace{-8pt}
\label{fig:scenario1}
\end{figure}

Suspicious of the association of a known disease named Corpus-Callosum-Agenesis (CCA) with the distribution of neural pathways at the intersection of the CC and CST bundles, Josh continues to customize the streamtube representation by mapping fractional
anisotropy (FA) to tube radius along each CST fiber since he is interested in the FA changes of CST at the intersection, and encoding depth values of CC fibers to colors so that he can easily discern the genu and splenium fibers in the CC bundle along the depth dimension in the coronal view. Finally, josh also wants to highlight the IFO fibers preferably represented with ribbons. Since the IFO bundle is roughly perpendicular to the CST bundle, he likes to take it as a reference as well. To achieve this task, Josh wrote the final script after error corrections as follows and got the result in the visualization view as shown in Figure~\ref{fig:scenario1}.
\begin{Verbatim}[frame=single]
LOAD "/home/josh/braindti.data"
SELECT "CC,CST,IFO"
UPDATE size BY FA IN "CST"
UPDATE depth BY color IN "CC"
UPDATE shape BY ribbon IN "IFO"
\end{Verbatim}

\subsection{Scenario 2: examining ROIs}
It is quite common that neurologists tend to examine particular regions of interest (ROIs) rather than the whole brain when using DTI visualizations. In this task, Josh is only interested in all fibers within the temporal lobe area that belong to the CG
bundle and CST fibers in the parietal lobe area that have average linear anisotropy (LA) value no larger than a threshold to
be determined. The SELECT command with relative spatial operations using the anatomical planes enables Josh to precisely
reach the ROIs he desires. 

\begin{figure}[!htb]
\centering
\includegraphics[width=8cm,height=6.5cm]{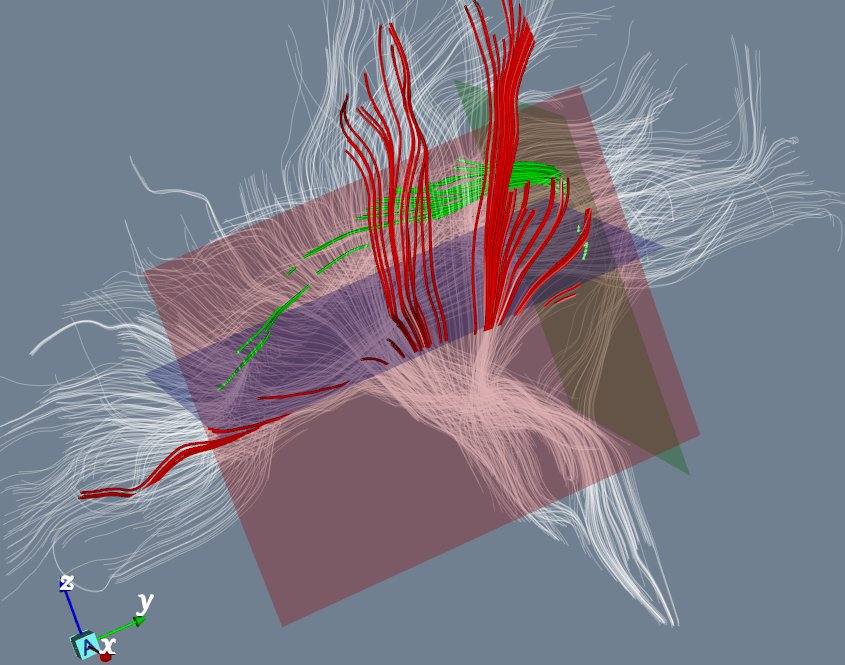}
\caption{Screenshot of the result after running the {\Tool} program that examines ROIs in scenario 2.}
\vspace{-8pt}
\label{fig:scenario2}
\end{figure}

To start with, Josh firstly aims to filter fiber tracts outside the temporal and parietal area by adjusting
the three cutting planes with relative movements and then starts trying to reach the exact target fiber tracts using both fiber bundle filters and conditional expression related to LA. With respect to the LA threshold undecided, Josh initially begins with
an estimate and then keeps refining until he gets the accurate selection of target fibers. In the end, he has a workable script
written in {\Tool} as follows. As a result, Figure~\ref{fig:scenario2} shows the ROIs that Josh programs for.
\begin{Verbatim}[frame=single]
LOAD "/home/josh/braindti.data"
SELECT "axial +63.35"
SELECT "sagittal +71"
SELECT "coronal -48.5"
SELECT "sagittal -0.25"
SELECT "axial +7.2"
SELECT "CG"
SELECT "LA <= 0.275" IN "CST"
\end{Verbatim}

\subsection{Scenario 3: calculating metrics}
Beyond visual examinations, neurologists often request quantitative investigations of their DTI models as well.
In this scenario, Josh attempts to check the white matter integrity in his brain model due to the limited reliability of
DTI tractography. 

\begin{figure}[htb]
\centering
\includegraphics[width=8cm,height=7cm]{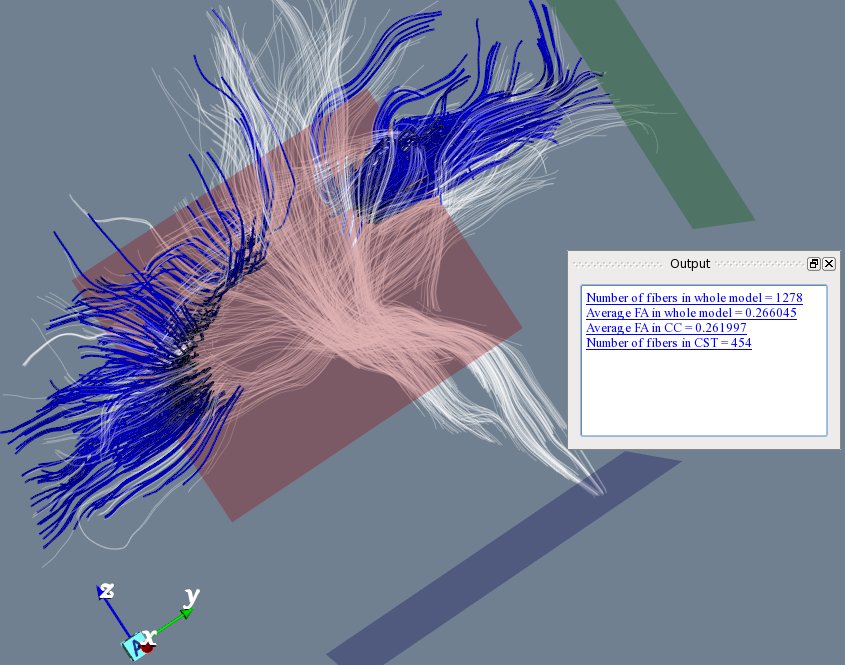}
\caption{Screenshot of the running result of {\Tool} script written for an end-user task in scenario 3.}
\vspace{-6pt}
\label{fig:scenario3}
\end{figure}

For a rough estimation of the integrity, he uses the CALCULATE command to retrieve the size, in
terms of the number of fibers, and average FA of both the whole brain and representative bundles. Further, with the
average FA he has requested before, Josh goes further to make use of it to kick out CST fibers with average FA below the
bundle-wise average. Josh writes the following script and obtains what he needs.
\begin{Verbatim}[frame=single]
LOAD "/home/josh/braindti.data"
SELECT "ALL"
CALCULATE NumFibers
CALCULATE AvgFA
cstFAavg = CALCULATE AvgFA in "CC"
CALCULATE NumFibers in "CST"
UPDATE RESET IN "ALL"
SELECT "FA >= cstFAavg IN "CC"
\end{Verbatim}
Figure~\ref{fig:scenario3} shows both the numerical values computed and the updated visualization using one of the
values through variable reference.

\section{Discussion}
\label{sec:discussion}
Since our language addresses scientific visualizations and targets non-programmer users, it is designed to be fully
declarative with flat control structure. While these two design features make the language easy to use for domain users,
it can cause difficulties in debugging the script since many low-level computations and control logics behind the
high-level syntax are hidden for the users. In order to minimize such drawbacks of the current {\Tool} design, the
script interpreter has been developed to strictly check each current statement and stop further executions of the script
once current return signals abnormal behaviours such as importing invalid data input and referring to unknown variables.

In addition, regarding the execution mode, the current implementation of {\Tool} does not follow a real-time interactive
running mode by which the visualization is updated once the script changes. Instead, the programming interface requires
a separate user interaction, such as clicking a button or pressing a key, for running the present script. This
design is for the interface simplicity and lower computational performance requirement, although
a programming environment with otherwise real-time update is easy to implement.

While at the prototype stage, {\Tool} is still under active development with an intention to add more useful features to
our visualization language for the purpose of better user experience and more powerful language expressiveness from
end-user prospectives in scientific domains. Some of the promising features are outlined as follows, which are also in our future plan for {\Tool} design and development.

\textbf{Concurrent multiple-model exploration:} While exploring more than one DTI models in order, i.e. switching data input
from one to another using the LOAD command, has been supported, concurrent exploration of multiple models has not yet.
However, requirements for doing so do exist among our end users. As an example, one typical case is to
examine two brain models in which one is known as normal and another suspicious of a brain disease. This is not rarely
seen in clinical practice since the side-by-side comparison is helpful for efficient recognition of cerebral anomalies
or simply finding structural differences. Corresponding {\Tool} commands and related other type of symbols can be
extended for such concurrent explorations. Among other changes, the evaluation of LOAD statement result to an identifier (a handle for instance) can be utilized to identify a specific model out of multiple ones simultaneously explored.

\textbf{Visual-aids for programming:} Even though programming provides controls more precise than visual interaction in
many cases, such as moving the axial plane by $22.5cm$, in a 3D visualization environment, under some occasions it is
hard to describe exploratory steps in the visualization. For example, with only an unverified picture in mind about the
outline of a fiber bundle that features human brain suffering a particular type of disease, a neurologist would like to
check if a given brain is afflicted with such disease by looking for any fiber bundles characteristic of the outline.
In this context, a visual aid that allows the user to sketch the outline for matching target fiber bundles can be fairly
effective while describing such outline in script is pretty difficult or at least far from being intuitive. Such visual
aids can be integrated into {\Tool}, with which users are enabled to designate a semantic term in the language
through visual sketching or drawing.

\textbf{Improved usability:} Although {\Tool} has been designed to be fully declarative and many features have
been developed expressly for maximal usability, such as flat control structure and consistent syntax
pattern, the usability of the overall programming environment can be further improved from two aspects. First of
all, apart from a help window showing all symbols and syntactical details which has already been implemented, context
-aware automatic word completion can be built into the script editor so that users would not need remember language
keywords. Also, statement templates can be provided in the interface so that users can program a statement simply by
filling blanks followed by clicking a button to confirm (then the statement will be added into the editor). Secondly,
instead of only displaying error message after execution, highlighting error-prone words when they
are being typed is an additional editor feature. 

\section{Conclusion}
\label{sec:conclusion}
We presented a visualization language for exploring 3D DTI visualizations and described the design principles and
language features of it derived from end-user descriptions about how to customize and explore such visualizations. We have
already developed some functions and features carefully selected of the proposed language, {\Tool}, and described the elements of the language. A primary design goal with {\Tool} is to initiate a scientific visualization language that is non-programmer oriented especially for domain scientists who have no any programming experience and skill to create and explore in their own visualizations. For this purpose, we emphasized design features of {\Tool} that particularly aim at our design goals.

We have also described representative usage scenarios of {\Tool} apart from many example scripts written in the language
before presenting its main content. These scenarios show that our new language is appealing to domain users and it is
promising to further develop the prototype towards a more capable and usable language for exploring more scientific visualizations.

While the development of our language as a whole is still at its early stage, the language core has already been
implemented and more features are being extended on top of current design. Among many possible directions to follow up,
we briefly discussed two main prospective features to follow up. By {\Tool} we have
presented a new approach, i.e. the end-user programming approach, to exploring DTI visualizations in 3D environment. This approach, as a complement to visual interactive environments, has a good potential to help narrow down the gap between visualization designers and end users with respect to the understanding of their underlying domain-specific data sets.

\section{Acknowledgements}
The authors are grateful to medical experts in DTI for their participation in our experimental studies, from which
the design features and principles of our visual programming language were abstracted. We would also like to thank their valuable comments for further enhancements of {\Tool} language when using it as end users.

\bibliographystyle{abbrv}
\bibliography{paper_arXiv}

\end{document}